 \documentstyle[twocolumn,aps,amssymb,amsmath,epsf,floats]{revtex} 
\begin{document}
\draft
\font\bfmit=cmmib10
\title{Anomalous Heavy-Fermion and Ordered States in the Filled-Skutterudite PrFe$_4$P$_{12}$}
 \twocolumn[ 
 \hsize\textwidth\columnwidth\hsize\csname@twocolumnfalse\endcsname 
\author{Y. Aoki, T. Namiki, T. D. Matsuda, K. Abe, H. Sugawara, and H. Sato} 
\address{Department of Physics, Tokyo Metropolitan University, Hachioji, Tokyo 192-0397, Japan} 

\date{\today}
\maketitle

\begin{abstract}
Specific heat and magnetization measurements have been performed on high-quality single crystals of filled-skutterudite PrFe$_4$P$_{12}$ in order to study the high-field heavy fermion state (HFS) and low-field ordered state (ODS).
From a broad hump observed in $C/T$ vs $T$ in HFS for magnetic fields applied along the $\langle 100 \rangle$ direction, the Kondo temperature of $\sim9$ K and the existence of ferromagnetic Pr-Pr interactions are deduced.
The $^{141}$Pr nuclear Schottky contribution, which works as a {\it highly-sensitive on-site probe} for the Pr magnetic moment, sets an upper bound for the ordered moment as $\sim0.03 \mu_B/$Pr-ion.
This fact strongly indicates that the primary order parameter in the ODS is nonmagnetic and most probably of quadrupolar origin, combined with other experimental facts.
Significantly suppressed heavy-fermion behavior in the ODS suggests a possibility that the quadrupolar degrees of freedom is essential for the heavy quasiparticle band formation in the HFS.
Possible crystalline-electric-field level schemes estimated from the anisotropy in the magnetization are consistent with this conjecture.
\end{abstract}
 ]
%


\narrowtext
\section{introduction} 

Studies of U-based intermetallic systems have revealed that the nature of strongly correlated electron states based on $f^2$ configuration can be qualitatively different from those based on $f^1$ configuration.
Such an example is possible quadrupolar Kondo coupling of a non-Kramers doublet ground state to conduction electrons in cubic symmetry \cite{Co87,Co98}.
This coupling, which may be viewed as a screening process of the fluctuating quadrupole moment on the ion by the conduction electrons, can lead to a non-Fermi-liquid (NFL) behavior if the coupling has symmetric two channels \cite{Ko99}.
This unconventional scenario has been applied to explain NFL behaviors observed in some cubic U compounds \cite{Se91,Al96,Am94}.
However, doubts have been thrown on the realization of 5$f^2$ configuration or a non-Kramers doublet ground state \cite{Ra94,Da95}.
The required substitution with non-magnetic elements for realization of NFL behavior allows other possible explanations to be proposed instead \cite{Mi97,Ca98}.

In this respect, the exploration of Pr-based correlated-electron systems is highly required because Pr$^{3+}$ (4$f^2$) is the most direct analogue to U$^{4+}$.
Compared to 5$f$ systems, smaller spatial extent of the 4$f$ orbitals usually causes more localized character with less valence fluctuation.
There is little debate regarding the existence of crystalline-electric-field (CEF) levels and, therefore, comparison with the aforementioned theoretical models becomes more reliable.
Although resulting weaker hybridizations of the 4$f$ electron might lead to a smaller energy scale (Kondo temperature $T_K$), which characterizes the renormalized properties developing at low temperatures, experimental observations are possible in limited systems. 
Such compounds reported to date are PrInAg$_2$ and PrFe$_4$P$_{12}$.
The former is characterized by the enhanced Sommerfeld coefficient $\gamma \sim 6.5$ J/K$^2$\ mol \cite{Ya96}.
Unusual $+\ln T$ dependence of electrical resistivity $\rho$ \cite{Mi00} might indicate NFL nature of the ground state in this compound.

In the present study, we focus on PrFe$_4$P$_{12}$, which exhibits a heavy-fermion state (HFS) in applied fields where an ordered state (ODS) is suppressed (see Fig. \ref{fig:HT} for a field-vs-temperature $H-T$ phase diagram).
The {\it Fermi-liquid} (FL) nature of HFS is evidenced by the enormous $\gamma$ value ($\simeq 1.4$ J/K$^2$\ mol in $\mu_0H=6$ T$//\langle110\rangle$ \cite{Ma00}), which satisfies the Kadowaki-Woods relation \cite{Ka86} along with the $T^2$ coefficient of electrical resistivity $\rho(T)$ ($2.5\ \mu \Omega$\ cm/K$^2$ in 6 T), and the huge cyclotron mass ($m_c^* \simeq 70$ m$_0$) determined by de Haas-van Alphen (dHvA) studies \cite{Su01}. 
To our knowledge, PrFe$_4$P$_{12}$ is the one and only system in which such definitive evidence for the 4$f^2$-based FL heavy-fermion ground state has been obtained.
In this paper, we report possible evidence indicating that the HF state is formed via nonmagnetic (most likely quadrupolar) interactions.

\section{experimental}

Single crystals of PrFe$_4$P$_{12}$ were grown by tin-flux method described in Ref. \cite{To87}.
The observation of the dHvA oscillations \cite{Su01} ensures high-quality of the sample. 
Specific heat $C(H,T)$ for $H//\langle100\rangle$ was measured by a quasi-adiabatic heat pulse method described in Ref. \cite{Ao98} using a dilution refrigerator equipped with an 8-T superconducting magnet.
The temperature increment caused by each heat pulse is controlled to $\sim$2\% for the usual measurement and to $\sim$0.5\% in limited temperature ranges where the phase transition occurs.
The bulk magnetization $M(\mu_0H\le 5.5$\ T$,T\ge 2.0$\ K$)$ was measured with a commercial SQUID magnetometer (Quantum Design).

\section{Results and discussions}

\begin{figure} 
 \centerline{\epsfxsize=7.5cm\epsfbox{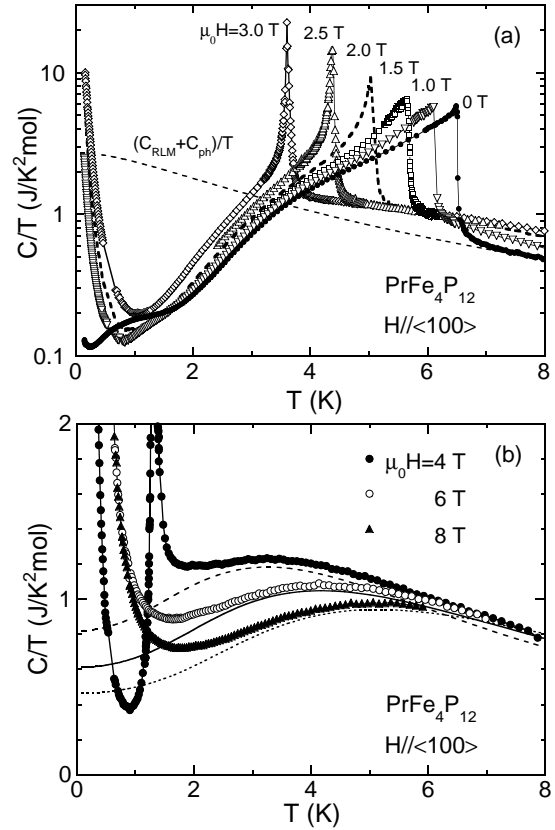}}
\caption{Total specific heat of PrFe$_4$P$_{12}$ in magnetic fields of (a) $\mu_0H \le 3$ T and (b) $\mu_0H\ge4$ T for $H//\langle100\rangle$. The thin curves represent the best model fittings of the HFS data using RLM (see text).}
\label{fig:COS}
\end{figure}

\begin{figure} 
 \centerline{\epsfxsize=7.5cm\epsfbox{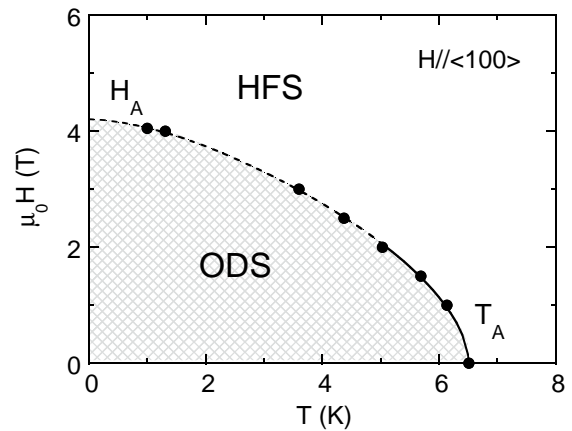}}
\caption{Magnetic field vs temperature phase diagram. The broken and solid curves represent first-order and second-order phase boundaries, respectively.}
\label{fig:HT}
\end{figure}

Figure \ref{fig:COS} shows the $C/T$-vs-$T$ data for $H//\langle100\rangle$.
In zero field, a clear anomaly indicating the second-order nature of the phase transition appears at $T_A=6.51$ K.
In the early stage, this was regarded as a signature of an antiferromagnetic transition since the magnetic susceptibility ($\chi$) drops below $T_A$ \cite{To87}.
With decreasing $T$ below $T_A$, $C/T$ decreases showing two weak shoulders at around 4 and 1 K.
A slight upturn below 0.2 K, which grows strongly in applied fields, is due to the nuclear Schottky contribution ($C_n$).
With increasing field, the anomaly at $T_A$ shifts to lower temperatures and changes its shape from second-order-like to first-order-like at $\sim 2$ T.
Actually, at temperatures very close to the phase transition in $\mu_0 H=2\sim 4$ T, Barkhausen-like heating is intermittently observed during heat pulse measurements.
The determined phase boundary between HFS and ODS is plotted in a $H-T$ phase diagram of Fig. \ref{fig:HT}.
In HFS above $\mu_0H=4$ T, there appears a broad hump at a temperature $T_{max}$, which shifts to higher temperatures with increasing field.
For $H//\langle110\rangle$ \cite{Ma00}, no such hump is observed up to 8 T although the overall $T$ dependence of $C/T$ is similar to that of the present data.
The hump structure and the enormously enhanced $C/T$ in the temperature range of $T\lesssim T_{max}$ cannot be explained by a simple Schottky model with a few separated levels.
Actually, $C/T$ does not show any exponential decrease with lowering temperature below $T_{max}$.
This fact indicates that, at least in the measured temperature range, "well-localized Pr ions with dominant CEF effect" does not provide an appropriate picture in PrFe$_4$P$_{12}$ and many body effects should be taken into account.

The measured specific heat $C$ can be expressed as $C_{el}+C_n+C_{ph}$.
For an estimation of the phonon contribution $C_{ph}$, we use $\beta T^3$ with $\beta=2.59 \times 10^{-4}$ J/K$^4$\ mol reported for LaFe$_4$P$_{12}$ \cite{Me84}.
Even at 8 K, $C_{ph}$ amounts to only 3\% of the total value of $C$ due to the huge contribution of $C_{el}$.

\begin{figure} 
 \centerline{\epsfxsize=7.5cm\epsfbox{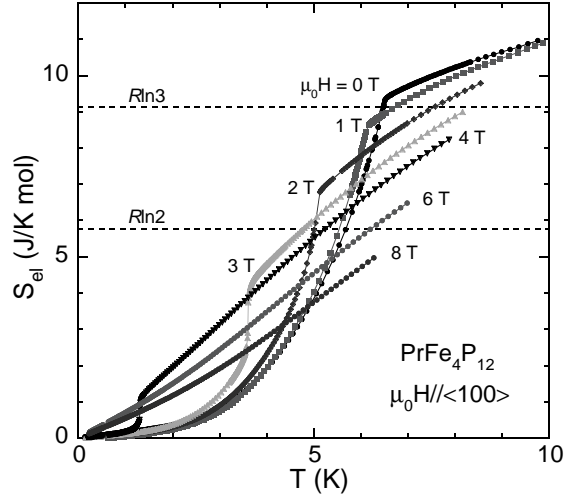}}
\caption{Temperature dependence of the electronic part of entropy $S_{el}$ calculated integrating the $C_{el}/T$ data. 
Because of the Barkhausen-like heating occurring at the first-order phase transition in $\mu_0 H=2\sim 4$ T, thus calculated $S_{el}$ above $T_A$ becomes lower than the true values.
By magnetization measurements using a Maxwell relation of $(\partial S/\partial (\mu_0H))_T=(\partial M/\partial T)_{\mu_0H}$, we have corrected the $S_{el}$ data in $T>T_A$.
In zero field, $S_{el}$ data for $T>T_A$ are nicely reproduced by the RLM model given by Eq.~(\ref{eq:RLMb}), which provides an asymptotic entropy of $S_{el}(T\to \infty)=R \ln (2^\delta)\simeq R \ln (6.5)$.}
\label{fig:S}
\end{figure}

The hump structure in $C/T$ vs $T$ is attributed to the low-energy thermal excitations in heavily renormalized quasiparticle bands formed around the Fermi level ($E_F$).
As a phenomenological model to describe the excitations, we employ the resonance level model (RLM) \cite{Sc75}, which assumes a Lorentzian density of states centered at $E_F$ with a $T$-independent width ($\Delta \sim T_K$).
Although this model is unrealistic, it has been applied successfully to describe the $C(T)$ data and to extract the characteristic energy scale on Ce-based HF compounds, e.g., CeRu$_2$Si$_2$ \cite{Meu91} and CeCu$_6$ \cite{Ed90}. 
We use the following slightly modified definition for the electronic specific heat, entropy, and magnetization:
\begin{subequations}
\begin{align}
C_{RLM} &=  \delta \cdot C_{SS}(T, \Delta, g\mu_BH_{eff}) \label{eq:RLMa} \\ 
S_{RLM} &=  \delta \cdot S_{SS}(T, \Delta, g\mu_BH_{eff}), \label{eq:RLMb} \\
M_{RLM} &=  \delta \cdot M_{SS}(T, \Delta, g\mu_B, H_{eff}), \label{eq:RLMc}
\end{align}
\end{subequations}
where the functions of $C_{SS}$, $S_{SS}$, and $M_{SS}$ are defined in Ref. \cite{Sc75} and $H_{eff}$ denotes an effective field acting on the electrons.
The prefactor $\delta$ is phenomenologically introduced considering that the entropy release associated with the hump structure is larger than $R \ln 2$ (see Fig.~\ref{fig:S}).
The connotation of $\delta$ will be discussed below.

\begin{figure} 
 \centerline{\epsfxsize=7.5cm\epsfbox{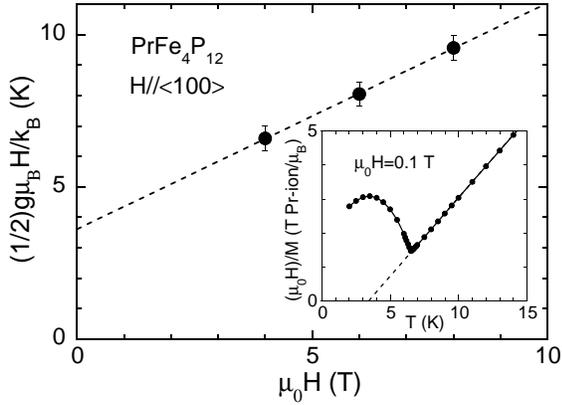}}
\caption{Magnetic field dependence of the effective Zeeman energy determined by the fitting shown in Fig. 1(b). The inset shows the $T$ dependence of inverse magnetic susceptibility.}
\label{fig:Zeeman}
\end{figure}

In zero field ($H_{eff}=0$), experimental values of $C_{el}/T=0.465$ J/K$^2$\ mol and electronic entropy $S_{el}=10.2$ J/K\ mol at $T=8$ K (see Fig.~\ref{fig:S}) uniquely lead to $\Delta=8.7$ K and $\delta=2.7$.
Model calculation of $C_{el}/T$ using these values is drawn in Fig. \ref{fig:COS}(a) by a broken thin curve.
The model curve, which satisfies the entropy balance with the zero-field data, approaches 2.7 J/K$^2$\ mol as $T\to 0$.
For the high-field data shown in Fig. \ref{fig:COS}(b), the only one fitting parameter is $g\mu_BH_{eff}$, which is determined so that $T_{max}$ can be reproduced.
Thus determined model curves shown in Fig. \ref{fig:COS}(b) demonstrate that the hump structure is described fairly well by Eq.~(\ref{eq:RLMa}).
The resulting effective Zeeman energy $(1/2)g\mu_BH_{eff}$ as a function of $H$ falls on a straight line as shown in Fig. \ref{fig:Zeeman}.
Assuming the effective $g$-factor to be $H$ independent between 4 and 8 T, $g=2.2$ is obtained from the slope in Fig. \ref{fig:Zeeman}.
The absence of the hump structure in $C/T$ for $H//\langle110\rangle$ up to 8 T \cite{Ma00} is due to smaller $g$ for this geometry, in accordance with the anisotropy in $M$ as will be discussed below.
The positive intercept in Fig. \ref{fig:Zeeman} indicates ferromagnetic interactions among the Pr ions with an energy scale of $\sim3.6$ K, which is consistent with the positive Weiss temperature $\theta_p=+3.5$ K determined from the magnetic susceptibility $\chi \equiv M/(\mu_0H)$ ($\mu_0H=0.1$\ T) data below 20 K (inset of Fig. \ref{fig:Zeeman}).
This feature contrasts strikingly with the negative $\theta_p$'s observed in Ce-based Kondo lattice compounds, where $|\theta_p|$ provides a rough estimate of $T_K$.
We also compare $M(T)$ measured in $\mu_0H=5.5$ T with $M_{RLM}(T)$ given by Eq.~(\ref{eq:RLMc}).
At 2.0 K, $M$ and $M_{RLM}$ are 1.83 and 1.58\ $\mu_B/$Pr-ion, respectively.
Except the slightly smaller value of $M_{RLM}$, the general feature of gradually decreasing $M$ with increasing $T$ is well reproduced (not shown).

The observed nuclear Schottky contribution $C_n$ is mostly caused by Pr nuclei (nuclear spin $I=5/2$ for $^{141}$Pr with the natural abundance of 100\%).
The contribution from Fe and P nuclei is negligibly small.
The nuclear magnetic moment averaged over the existing isotopes ($\overline{I(I+1)g_N^2}$) is $5.3\times 10^{-4}$ and 3.84 for Fe and P, respectively.
Using these values, the contribution from Fe and P nuclei can be estimated as $C_n=1.7\times 10^{-5} (\mu_0H_{hf}/T)^2$ J/K\ mol, which amounts to only 0.4\% of the observed $C_n$ in 8 T assuming $\mu_0H_{hf}\simeq 8$ T.

The observed Pr nuclear Schottky contribution is largely enhanced.
This is due to the strong intra-site hyperfine coupling between the nucleus and 4{\it f}-electrons on the same Pr ion.
The Hamiltonian for the Pr nucleus can be written as \cite{Su78}
\begin{equation}
\label{eq:H}{\cal H}_n=a'I_z + P(I_z^2-I(I+1)/3),
\end{equation}
where the $z$ axis lies in the direction of $\langle{\bf J}\rangle$ for each Pr ion.
Here $a'$ ($=A \langle J_z \rangle$) and $P$ ($=B \langle J_z^2-J(J+1)/3 \rangle$) are the magnetic dipole hyperfine interaction parameter and the electric quadrupole coupling constant, respectively.
For each magnetic field, the values of $a'$ and $P$ can be determined by fitting the observed $C_n(T)$ data to the model given by Eq.~(\ref{eq:H}).
Consequently, if the coupling constant $A$ is known, $\langle J_z \rangle$ will be obtained.
In the following data analysis, we replace $\langle J_z \rangle$ by the site-averaged quantity $(\overline{|\langle J_z \rangle|^2})^{1/2}$ since the Pr ions are most probably separated into inequivalent sites in ODS and the measured specific heat provides the site-averaged quantity.
Using this, we define the site-averaged magnitude of the Pr magnetic moment $m_{Pr} \equiv g_J (\overline{|\langle J_z \rangle|^2})^{1/2}$.

We experimentally determine the coupling constant $A$ in HFS, where $\langle{\bf J}\rangle$ is polarized along the applied field for all Pr ions.
In $\mu_0H=5.5$ T, $M(T \to 0)=g_J \langle J_z \rangle=1.89\ \mu_B$/Pr-ion and $a'$ is obtained to be 0.122 K.
Thus determined $A=0.052$ K agrees well with reported theoretical values \cite{Ko61,Bl63}.

\begin{figure} 
 \centerline{\epsfxsize=7.5cm\epsfbox{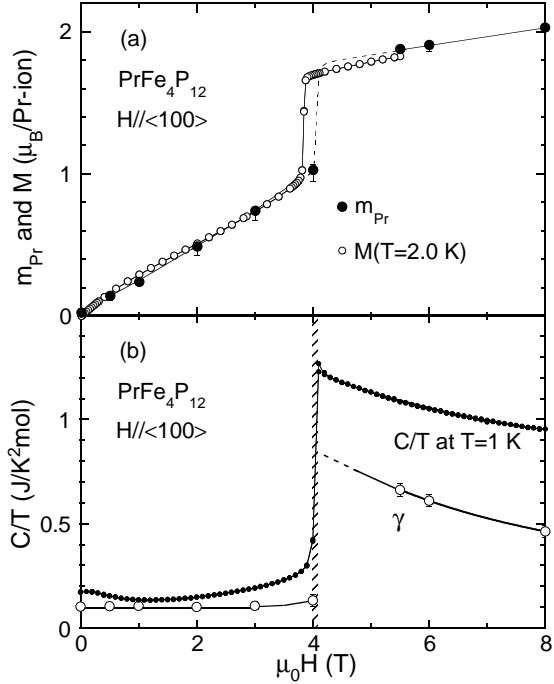}}
\caption{(a) Magnetic field dependences of the site-averaged magnitude of the Pr magnetic moment $m_{Pr}$ and bulk magnetization $M$($T=2.0$ K). (b) $\gamma \equiv C_{el}/T|_{T\to 0}$ estimated by the fitting below 2 K (see text). $C/T$ vs $\mu_0H$ measured at 1 K is shown for comparison.}
\label{fig:a}
\end{figure}

In order to separate $C_n$ and $C_{el}$, we use Eq.~(\ref{eq:RLMa}) in HFS and $\gamma T+\alpha T^n$ in ODS for $C_{el}$.
The $H$ dependences of $m_{Pr}$ and $\gamma$, obtained by fitting $C(T)$ data below 2 K, are shown in Fig. \ref{fig:a}.
The resulting exponent $n$ ranges from 4.1 to 4.3.
In HFS, $P$ is determined to be $(7 \pm 2)\times 10^{-3}$ K; for the precise determination of $P$ in low fields, the measurement should be extended further below 0.1 K.

A remarkable feature in Fig. \ref{fig:a}(b) is that $\gamma$ is substantially reduced ($\simeq 0.1$ J/K$^2$\ mol) and almost $H$ independent in ODS, and it rises sharply to $\sim 1$ J/K$^2$\ mol at $H_A$. 
Field-scan data of $C/T$ measured at 1 K (shown in Fig. \ref{fig:a}(b) for comparison) also reflects this feature; the difference between the $C/T$ and $\gamma$ data comes mainly from the nuclear contribution in HFS and the magnetic excitation approximated as $\alpha T^n$ in ODS.
This observation evidently shows that the HF behavior is significantly suppressed in ODS.

In Fig. \ref{fig:a}(a), $m_{Pr}$ is compared with $M(H)$ data measured at 2.0 K.
A small difference in $H_A$ is due to the temperature dependence of $H_A$ (see Fig. \ref{fig:HT}).
Agreement between the two sets of data is quite good even in ODS taking into account the weak temperature dependence of $M$; e.g., $M(T \to 0)$ is only 1\% smaller than $M(T=2$\ K$)$ in $\mu_0H=3$ T.
This agreement clearly shows that the reduced $M$ in ODS is mainly due to the shrinkage of each Pr magnetic moment itself, not due to cancellation among the Pr magnetic moments as usually observed in classical antiferromagnets.
For example, $M(H)$ curve of HoGa$_2$ shows complex field-induced metamagnetic phase transitions, though the Ho nuclear specific heat (i.e., the magnitude of each Ho magnetic moment) is of $H$ independence, reflecting that the metamagnetic behavior in $M(H)$ is due to the change in the Ho moment cancellation depending on the field-induced magnetic structures \cite{Ao00}.

\begin{figure} 
 \centerline{\epsfxsize=7.5cm\epsfbox{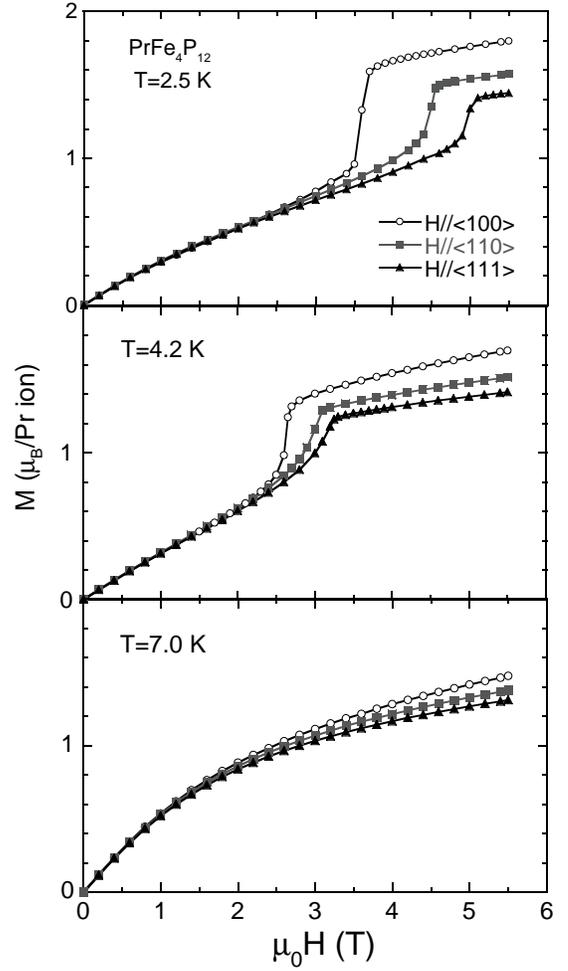}}
\caption{Magnetization curves measured at 2.5, 4.2 and 7 K in external fields along the main symmetry directions. The observed magnetic anisotropy of $M(H//\langle100\rangle)>M(H//\langle110\rangle)>M(H//\langle111\rangle)$ is the feature of the non-ordered state.}
\label{fig:MH}
\end{figure}
As shown in Fig. \ref{fig:a}(a), $m_{Pr}$ decreases significantly as $H\to 0$.
The extremely small $C_n\simeq 6.4 \times 10^{-5}/T^2$ J/K\ mol observed in zero field sets an upper bound for the ordered $m_{Pr}$, i.e.,$\le 0.03 \mu_B$/Pr-ion.
The extremely small $m_{Pr}$ along with the large entropy decrease below $T_A$ as shown in Fig. \ref{fig:S} strongly suggests that the primary order parameter in ODS is not the Pr magnetic moment but of non-magnetic origin.
This is consistent with the absence of magnetic Bragg peaks at 1.5 K in a powder neutron diffraction study \cite{Ke00}, in which typical sensitivity is of the order of 0.1 $\mu_B$/Pr-ion.

\begin{figure} 
 \centerline{\epsfxsize=7.5cm\epsfbox{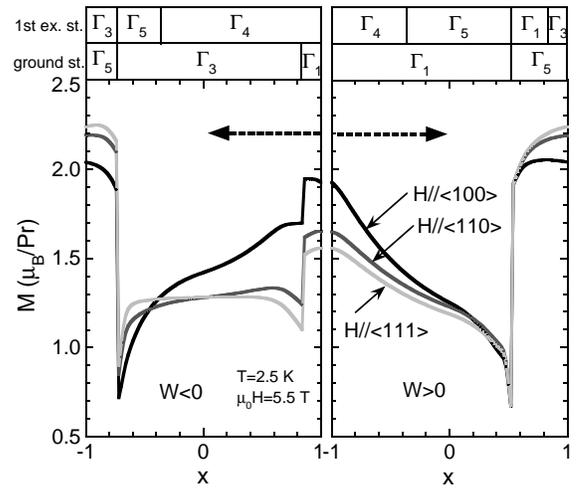}}
\caption{$x$ dependence of the calculated magnetization for a single-site Pr ion at 2.5 K in a magnetic field of 5.5 T. For each $x$, the absolute value of $W$ is determined so that $S_{el}(T=8$ K$)=10.2$ J/K mol is satisfied. No molecular field for the dipole or multipole Pr-Pr interactions is taken into account. Broken arrows indicate the parameter region where the magnetization anisotropy is qualitatively consistent with the experimental data. The upper panel shows the ground and first excited CEF states.}
\label{fig:Mcalc}
\end{figure}

In the non-ordered state, it is found that the low-temperature properties cannot be described quantitatively in a well-localized Pr-ion picture.
For example, the specific heat cannot be described by a simple CEF model producing a Schottky anomaly.
It is expected, however, that some of the qualitative features may reflect the CEF effect playing a role in the background of the strong hybridization.
The $T_h$ site symmetry of the Pr$^{3+}$ ions with no fourfold rotation axis leaves a non-vanishing $(O^2_6-O^6_6)$ term in the CEF Hamiltonian \cite{Ta00}, along with the usual cubic terms of $(O^0_4+5O^4_4)$ and $(O^0_6-21O^4_6)$.
If the $(O^2_6-O^6_6)$ term is dominating, one of the two magnetic triplets will be stabilized as the CEF ground state and consequently the anisotropy in $M$ will be $M(H//\langle111\rangle)>M(H//\langle110\rangle)>M(H//\langle100\rangle)$ at low temperatures.
Magnetization curves measured in applied fields along the main symmetry directions are shown in Fig.~\ref{fig:MH}.
Since $M(H//\langle100\rangle)>M(H//\langle110\rangle)>M(H//\langle111\rangle)$ is observed in the non-ordered state, the above possibility is ruled out.
In the following analysis, therefore, we assume that the $(O^2_6-O^6_6)$ term is negligibly small and we use the $O_h$ point-group notation for CEF levels described in Lea, Leask and Wolf (LLW) \cite{Le62} (see Ref. \cite{Ta00} for the $O_h \leftrightarrow T_h$ correlation table).
In this notation, the two CEF parameters are represented by $W$ and $x$.
We calculated magnetization for a single-site Pr ion at 2.5 K in a magnetic field of 5.5 T in order to compare with the experimental data.
The results are shown in Fig.~\ref{fig:Mcalc} as a function of $x$.
For each $x$, the absolute value of $W$ is determined so that $S_{el}(T=8$ K$)=10.2$ J/K mol is satisfied.
No molecular field for the dipole or multipole Pr-Pr interactions is taken into account in the calculation.
The observed anisotropy in $M$ is qualitatively reproduced when $0 \lesssim x \le 1$ for $W<0$ or $-1\le x \lesssim0$ for $W>0$.
In these parameter regions, the CEF ground state is a $\Gamma_1$ singlet or a $\Gamma_3$ non-Kramers doublet; in either case the ground state is non-magnetic, in accordance with the non-magnetic nature of ODS.
A $\Gamma_5$ triplet ground state is highly unlikely.

The magnetic susceptibility $\chi$ in ODS has quite large values, as readily recognized in both the inset of Fig. \ref{fig:Zeeman} and the magnetization curve shown in Fig.~\ref{fig:a}(a).
If the observed $\chi$ in ODS were attributed to itinerant electrons with $\gamma \simeq 0.1$ J/K$^2$\ mol, an unreasonably large effective moment $\mu_{eff}=15\ \mu_B$/carrier would be needed, indicating that the observed $\chi$ in ODS is attributed to the well-localized 4{\it f} electrons, of which the $c$-$f$ hybridization is substantially suppressed.
Attributing $\chi \simeq0.4\ \mu_B$/T\ Pr-ion for $T \to 0$ in ODS to a Van Vleck susceptibility, following possible CEF level schemes are deduced.
(i) In the case of the $\Gamma_1$ ground state, a $\Gamma_4$ state is located at $E(\Gamma_4) \simeq 13$ K above the ground state, taking into account the $\Gamma_1$ ground state having only one non-zero matrix element of $J_z$ with the $\Gamma_4$ state.
Since the experimental $S_{el}(T=8$\ K$)=10.2$ J/K\ mol shown in Fig.~\ref{fig:S} denies the possibility that a $\Gamma_5$ state is located below $\sim 13$ K as the first excited state, one can conclude that the $\Gamma_4$ state is the first excited state and the $\Gamma_5$ state should be located above it.
Calculated entropy at 8 K in zero field (in the non-ordered state) neglecting the $\Gamma_5$ state and the other higher levels is 8.9 J/K\ mol, which is not far from the experimental value.
(ii) In the case of the $\Gamma_3$ ground state, $\Gamma_3^{(1)}$ referring to $\sqrt{7/24}(|4\rangle+|-4\rangle)-\sqrt{5/12}|0\rangle$ ($|M\rangle$ represents an eigenstate of $J_z$) has only one non-zero matrix element of $J_z$ with $\Gamma_4$ state and so does $\Gamma_3^{(2)}$ referring to $\sqrt{1/2}(|2\rangle+|-2\rangle)$ with $\Gamma_5$ state.
Therefore, with the similar consideration as in the case (i), one can conclude that the first excited state is a $\Gamma_4$ state, which is located at $E(\Gamma_4) \simeq 18$ K if it can be assumed that all the Pr ions have the $\Gamma_3^{(1)}$ ground state in ODS.
Using the value of $E(\Gamma_4)$, calculated entropy at 8 K in zero field (in the non-ordered state) neglecting the $\Gamma_5$ state and the other higher levels is 9.6 J/K\ mol, which is also not far from the experimental value.
From the rough discussion mentioned above based on the present specific-heat and magnetization results, we propose that the possible CEF level schemes are $\Gamma_1-\Gamma_4(\sim$13~K$)$ and $\Gamma_3-\Gamma_4(\sim$18~K$)$.
Noted that the energy level of the first excited state should be considered as a rough estimation since we have made some assumptions in the discussion, e.g., the modification due to the ordering of the wave functions and the energy levels of the CEF levels to be small enough.
In order to check our proposal, inelastic neutron scattering measurement is planed.

Ultrasonic study reported by Nakanishi {\it et al.} \cite{Na01} demonstrates that the elastic constants of $c_{11}$ and $c_{11}-c_{12}$ shows softening below $\sim20$ K down to $T_A$ while $c_{44}$ does not.
From this fact, they claim that the CEF level scheme is $\Gamma_3-\Gamma_4$, which is the same as one of our proposed models.
Below $T_A$, both $c_{11}$ and $c_{11}-c_{12}$ start to increase.
From these observations, they proposed that the order parameter in ODS would be an antiferro quadrupole ordering with $\Gamma_3$ symmetry, where the twofold degeneracy of the $\Gamma_3$ ground state is lifted.
If this is the case, $S_{el}(T=T_A)$ being larger than $R\ln2$ indicates that the excited $\Gamma_4$ state, which also has the quadrupole moments, is also relevant to the ordering.

In the case of $\Gamma_1-\Gamma_4$ level scheme, which cannot be ruled out by our present study, the $\Gamma_1$ ground state itself has no degrees of freedom.
However, since the energy separation between the two levels is comparable to the estimated value of $T_K$, one can speculate that the $\Gamma_4$ excited state contributes much to the renormalized physical properties at low temperatures.
Due to the quadrupole degrees of freedom relevant to the $\Gamma_4$ state, a quadrupole ordering could take place also in this case.

In ODS, where the quadrupolar degeneracy is most probably lifted, the strong suppression of the HF behavior is observed as shown in Fig.~\ref{fig:a}(b).
This observation points to a possibility that the fluctuation of the quadrupole moments in the non-ordered state is playing an essential role for the heavy-fermion state formation.
Figure~\ref{fig:a}(a) shows that $m_{Pr}$ develops monotonously with increasing magnetic field and approaches as large as 60\% of the HFS value at $H_A$.
The insensitivity of the suppression of the HF behavior to the field induced $m_{Pr}$ in ODS implies that the suppression of the HF behavior is not due to any magnetic sources, probably consistent with our conjecture.

Any sign of possible NFL behaviors, which could be caused by the quadrupolar Kondo effect, have not been observed in PrFe$_4$P$_{12}$, contrasting remarkably with PrAgIn$_2$.
The instability of NFL state against FL state has been studied theoretically in some model cases \cite{NFLFL}, although comparison with the real systems are still insufficient.
In PrFe$_4$P$_{12}$, largely-enhanced negative thermopower below 50 K \cite{Sa00} suggests the predominant hybridization process to be between $f^2$ and $f^3$ configurations \cite{TEP}.
The virtual $f^3$ configuration is expected to have a quartet ground state since the CEF ground state of NdFe$_4$P$_{12}$ is a quartet \cite{To87}.
In this scheme, the {\it c-f} hybridization should have complicated multi-channels.
Another factor needed to be considered is the closely-located CEF excited state in the $f^2$ configuration in PrFe$_4$P$_{12}$.
In the case of the $\Gamma_3-\Gamma_4$ level scheme, the $\Gamma_4$ excited state might cause the instability of the NFL state relevant to the $\Gamma_3$ ground state, as demonstrated in a non-Kramers doublet and a singlet case \cite{NFLFL}.
For deeper understanding of the NFL-against-FL stability issue, theoretical study on the present case, which has not been done yet, is highly required.

A band structure calculation on LaFe$_4$P$_{12}$, which consistently explains observed angular dependence of the dHvA branches, indicates that the Fermi surface consists of a nearly spherical hole sheet (47th) and a multiply connected one (48th) \cite{Su00}.
The latter one is characterized by strong $c$-$f$ hybridizations.
Therefore, the heavily renormalized quasiparticle bands observed in PrFe$_4$P$_{12}$ are expected to be mostly attributed to the corresponding 48th sheet of LaFe$_4$P$_{12}$.
As pointed out by Harima \cite{Su00}, its nearly cubic shape may favor a commensurate 3-dimentional nesting with ${\bf q}=\{1,0,0\}$.
This conjecture is supported by some of the corresponding X-ray satellite peaks observed in ODS \cite{Iw00} and the carrier reduction suggested by jumps in the electrical resistivity and the Hall coefficient at $T_A$ \cite{Sa00}.
The lattice distortion (or the charge ordering) and the associated gap opening on some part of the corresponding Fermi sheet should be also taken into account to understand the nature of ODS.

\section{summary}

We have studied the heavy-fermion and the anomalous ordered states in filled-skutterudite PrFe$_4$P$_{12}$ by specific heat and magnetization measurements.
From a broad hump observed in $C/T$ vs $T$ for $H//\langle 100 \rangle$ in HFS, the Kondo temperature of $\sim9$ K and the existence of ferromagnetic Pr-Pr interactions are deduced.
Utilizing $^{141}$Pr nuclear Schottky contribution, an upper bound for the site-averaged Pr ordered magnetic moment is obtained to be $\sim0.03 \mu_B/$Pr-ion.
This fact strongly indicates that the primary order parameter in the ordered state is nonmagnetic, most probably of quadrupolar origin, as suggested by the ultrasound measurements.
Significantly suppressed heavy-fermion behavior in the ordered state suggests a possibility that the quadrupolar fluctuation of the Pr ions leads to the heavy quasiparticle band formation in the high-field heavy fermion state.
Possible crystalline-electric-field level schemes of the Pr$^{3+}$ ions estimated from the anisotropy in the magnetization and the electronic part of entropy are $\Gamma_1-\Gamma_4$ and $\Gamma_3-\Gamma_4$ [(ground state)-(first excited state)], both of which are consistent with the aforementioned conjecture.
The present study demonstrates that PrFe$_4$P$_{12}$ could be an unprecedented system, where the role of the quadrupole moments can be studied for the realization of the rare 4$f^2$-based heavy fermion state thanks to the existence of the ordered state, which is not available in PrInAg$_2$.

\acknowledgments

We thank A.~D\"onni, K.~Takegahara, H. Harima, O. Sakai, R. Shiina, M. Koga, K. Miyake and K. Ueda for useful discussions.
This work is supported partly by a Grant-in-Aid for Scientific Research from the Ministry of Education, Science and Culture and by the REIMEI Research Resources of Japan Atomic Energy Research Institute.

\end{document}